\documentclass[11pt,draftcls,onecolumn]{IEEEtran}
\usepackage{amsmath}
\usepackage{amsthm}
\usepackage{graphicx}
\usepackage{color}
\usepackage{subfigure}
\usepackage{cite}
\usepackage{url}

\usepackage{algorithm}
\usepackage{algorithmic}

\theoremstyle{plain}

\theoremstyle{definition}

\title{Chinese Restaurant Game - Part II: Applications to Wireless Networking, Cloud Computing, and Online Social Networking}
\author{
\IEEEauthorblockN{Chih-Yu Wang$^{1,2}$, Yan Chen$^1$, and K. J. Ray Liu$^1$}\\
\IEEEauthorblockA{
$^1$Department of Electrical and Computer Engineering, University of Maryland, College Park, MD 20742 USA\\
$^2$Graduate Institute of Communication Engineering, National Taiwan University, \\
Taipei, Taiwan
}
}
\begin{document}

\maketitle

\begin{abstract}
In Part I of this two-part paper \cite{wang2011crgpart1}, we proposed a new game, called Chinese restaurant game, to analyze the social learning problem with negative network externality. The best responses of agents in the Chinese restaurant game with imperfect signals are constructed through a recursive method, and the influence of both learning and network externality on the utilities of agents is studied. In Part II of this two-part paper, we illustrate three applications of Chinese restaurant game in wireless networking, cloud computing, and online social networking. For each application, we formulate the corresponding problem as a Chinese restaurant game and analyze how agents learn and make strategic decisions in the problem. The proposed method is compared with four common-sense methods in terms of agents' utilities and the overall system performance through simulations. We find that the proposed Chinese restaurant game theoretic approach indeed helps agents make better decisions and improves the overall system performance. Furthermore, agents with different decision orders have different advantages in terms of their utilities, which also verifies the conclusions drawn in Part I of this two-part paper.
\end{abstract}

\newpage

\section{Introduction}

The network externality, which describes the mutual influence among agents, plays an important role in numerous network-related applications. When the network externality is negative, i.e., the more agents make the same decision, the lower utilities they have in the network, agents tend to avoid making the same decision with others in order to maximize their utilities. This phenomenon has been observed in many applications in various research areas, such as dynamic spectrum accessing in cognitive radio networking, service selection in cloud computing, and deal selection on Groupon. Moreover, agents may not know the exact state of the network, such as the primary user's activity in the spectrum, the infrastructure of service platform, and qualities of products provided in deals. Therefore, they tend to learn such unknown information from some signals through measurements or actions from others in the network, which is called social learning. How these two effects, negative network externality and social learning, affect the decisions of agents in different network-related applications, is the main objective of this paper. Chinese Restaurant Game,  proposed in Part I of this two-part paper \cite{wang2011crgpart1}, provides a general framework for modeling strategic learning and decision processes in the social learning problem with negative network externality.

Here, we briefly describe the proposed Chinese restaurant game. We consider a Chinese restaurant with $K$ tables numbered $1,2,...,K$ and $N$ customers labeled with $1,2,...,N$. Each table has infinite seats, but may be in different size. We model the table sizes of a restaurant with two components: the restaurant state $\theta$ and the table size functions $\{R_1(\theta),R_2(\theta),...,R_K(\theta)\}$. The restaurant state $\theta$ is unknown to the customers. However, each customer $i$ has received a signal $s_i$ following a commonly known distribution $f(s|\theta)$. Therefore, customers may learn the state from the signals they collected. In the Chinese restaurant game, each customer sequentially requests for the table following the same order as their numbering. After customer $i$ made his request, he reveals both his decision and the signal he received. Let $x_i$ be the table requested by customer $i$. Then, customer $i$'s payoff is determined by a utility function $u_i(R_{x_i},n_{x_i})$, where $n_{x_i}$ is the number of customers choosing table $x_i$.

The best responses of rational customers that maximize their expected utilities can be recursively derived in Chinese restaurant game \cite{wang2011crgpart1}. We denote $\mathbf{n}=\{n_1,n_2,...,n_K\}$ as the numbers of customers on the $K$ tables, i.e., the grouping of customers in the restaurant at the end of the game. Let $\mathbf{n_i}=\{n_{i,1},n_{i,2},...,n_{i,K}\}$ be the grouping observed by customer $i$ when making his decision, and $\mathbf{h_i}=\{s_1,s_2,...,s_{i-1}\}$ be the history of revealed signals before customer $i$. The best response of customer $i$ can be written as
\begin{equation}\label{eqn34}
    x_i = BE_i(\mathbf{n_i},\mathbf{h_i},s_i) = \arg\max_j E[u_i(R_j(\theta),n_j)|\mathbf{n_i}, \mathbf{h_i}, s_i].
\end{equation}

Customers in the Chinese restaurant game need to estimate the current state $\theta$ with the signals they collected in order to make the right decisions. Let the prior probability of the state be $\mathbf{g_0} =\{g_{0,l}|g_{0,l}=Pr(\theta=l),~\forall l \in \Theta\}$. We denote customer $i$'s estimate on the current state as the belief being defined as $\mathbf{g_i} = \{g_{i,l}|g_{i,l}=Pr(\theta=l|\mathbf{h_i},s_i,g_0,f),~\forall l \in \Theta\} ~\forall i \in N$. Then, customer $i$ can construct his belief according to the Bayesian rule as follows:
\begin{equation}\label{eqn33}
    g_{i,l} = \frac{Pr(\mathbf{h_i},s_i|\theta=l)Pr(\theta=l)}{\sum_{w=1}^L {Prob(\mathbf{h_i},s_i|\theta=w)Pr(\theta=w)}} = \frac{g_{i-1,l}f(s_i|\theta=l)}{\sum_{w=1}^L {g_{i-1,w}f(s_i|\theta=w)}}.
\end{equation}

Besides the estimate on the state $\theta$, customer $i$ also needs to predict the decisions of subsequent customers due to the negative network externality effect. Let $m_{i,j}$ be the number of customers choosing table $j$ after customer $i$, including customer $i$ himself. As shown in \cite{wang2011crgpart1}, we have

\begin{eqnarray}\label{eqn39}
\!\!\!\!\!\!\!\!\!\!\!\!&&\!\!\!\!\!\!\!\!\!\!\!\!Pr(m_{i,j}=X|\mathbf{n_i},\mathbf{h_i},s_i,x_i,\theta=l) =\left\{
                                            \begin{array}{ll}
                                              Pr(m_{i+1,j}=X-1|\mathbf{n_i},\mathbf{h_i},s_i,x_i,\theta=l), & \hbox{$x_i=j$,} \\
                                              Pr(m_{i+1,j}=X|\mathbf{n_i},\mathbf{h_i},s_i,x_i,\theta=l), & \hbox{$x_i \neq j$,}
                                            \end{array}
                                          \right. \\
                                          \!\!\!\!\!\!\!\!\!\!\!\!&&\!\!\!\!\!\!\!\!\!\!\!\!=\!\!\!\left\{\!\!\!\!
                                               \begin{array}{ll}
                                                 \sum_{0 \leq u \leq K} \int_{s \in S_{i+1,u}(\mathbf{n_{i+1}},\mathbf{h_{i+1}})} \nonumber {Pr(m_{i+1,j}=X-1|\mathbf{n_{i+1}},\mathbf{h_{i+1}},s_{i+1}=s,x_{i+1}=u, \theta=l)} f(s|\theta=l) ds, &\!\!\! \hbox{$x_i=j$,} \\
                                                 \sum_{0 \leq u \leq K} \int_{s \in S_{i+1,u}(\mathbf{n_{i+1}},\mathbf{h_{i+1}})} {Pr(m_{i+1,j}=X|\mathbf{n_{i+1}},\mathbf{h_{i+1}},s_{i+1}=s,x_{i+1}=u, \theta=l)} f(s|\theta=l) ds, &\!\!\! \hbox{$x_i \neq j$,}
                                               \end{array}
                                             \right.
\end{eqnarray}
where $\mathbf{h_{i+1}}$ and $\mathbf{n_{i+1}}$ can be obtained using
\begin{equation}\label{eq_h_plus}
\mathbf{h_{i+1}}=\{h_{i},s_i\} \text{ and } \mathbf{n_{i+1}}=\{n_{i+1,1},...,n_{i+1,K}\},
\end{equation}
with
\begin{equation}\label{eq_n_plus}
n_{i+1,k}=\left\{
        \begin{array}{ll}
                 n_{i,k}+1, & \mbox{if $x_i=k$}, \\
                 n_{i,k}, & \mbox{otherwise}. \\
        \end{array}
        \right.
\end{equation}

Based on (\ref{eqn39}), $Pr(m_{i,j}=X|\mathbf{n_{i}},\mathbf{h_{i}},s_{i},x_{i},\theta=l)$ can be recursively calculated. Then, we can compute the expected utility $E[u_i(R_j(\theta),n_j)|\mathbf{n_i}, \mathbf{h_i}, s_i]$ and derive the best response function of customer $i$ as
\begin{equation}\label{eqn44}
    BE_i(\mathbf{n_i},\mathbf{h_i},s_i) = \arg\max_j \sum_{l \in \Theta} \sum_{x=0}^{N-i+1} g_{i,l}Pr(m_{i,j}=x|\mathbf{n_i},\mathbf{h_i},s_i,x_i=j,\theta=l)u_i(R_j(l),n_{i,j}+x).
\end{equation}

With the recursive form, the best response function of all customers can be obtained using backward induction. In Part I of this two-part paper \cite{wang2011crgpart1}, we generally discussed how customers make decisions in the Chinese restaurant game under different signal qualities and table size ratios. In Part II of this two-part paper, we would like to illustrate how Chinese restaurant game can be used in specific applications in different research areas. In Section \ref{sec_comp}, we first describe four strategies that we will compare our method with. Then, we will show the improvement from the proposed best response strategy on both the individual utility and the overall system efficiency in three applications: dynamic spectrum access in cognitive radio networking, storage service selection in cloud computing, and deal selection on Groupon in online social networking in Section \ref{sec_app1}, \ref{sec_app2}, and \ref{sec_app3}, respectively. For each application, we first formulate the problem using the Chinese restaurant game. Then, we compare the proposed best response strategy with other strategies in terms of agents' utilities and the overall system performance through simulations. Finally, we draw our conclusions in Section \ref{sec_con}.

\section{Strategies for Comparisons}\label{sec_comp}
We will compare our best response strategy with the following four strategies: random, signal, learning, and myopic strategies. In the random strategy, customers choose their strategies randomly and uniformly, i.e., all $K$ tables have equal probability of $\frac{1}{K}$ to be chosen under the random strategy. In the signal strategy, customers make their decisions purely based on their own signal regardless all information from other customers, including the revealed signals and their choices on tables. The objective of signal strategy is to choose the largest expected table size conditioning on his signal given by
\begin{equation}
    x^{signal}_i = \arg \max_x \sum_{l \in \Theta} Pr(s_i|\theta=l) R_x(l).
\end{equation}

The learning strategy is an extension of the signal strategy. Under this strategy, the customer learns the system state not only by his own signal but also by the signals revealed by the previous customers. Therefore, the learning strategy can be obtained as
\begin{equation}
    x^{learn}_i = \arg \max_x \sum_{l \in \Theta} g_{i,l} R_x(l),
\end{equation}
where $g_{i,l}=Pr(\theta = l| s_1,s_2,...,s_i, \mathbf{g_0})$ is the belief of the customer on the state.

Finally, the myopic strategy simulates the behavior of a myopic player. The objective of a customer under myopic strategy is maximizing his current utility, i.e., the customer makes the decision according to his own signal, all signals revealed by previous customers, and the current grouping $\mathbf{n_i}=\{n_{i,1},n_{i,2},...,n_{i,K}\}$ as follows,
\begin{equation}\label{eq_myopic}
    x^{myopic}_i= \arg \max_x \sum_{l \in \Theta} g_{i,l} u_i(R_x(l),n_{i,x}+1).
\end{equation}

From (\ref{eq_myopic}), we can see that the myopic strategy is similar to the proposed best response strategy except the Bayesian prediction of the subsequent customers' decisions. The performance of all these four strategies will be evaluated in all simulations in the following applications. They will be treated as the baseline of the system performance without fully rational behaviors of customers.

\section{Wireless Networking: Dynamic Spectrum Access in Cognitive Radio Network}\label{sec_app1}
Traditional dynamic spectrum access methods focus on identifying available spectrum through spectrum sensing. Cooperative spectrum sensing is a potential scheme to enhance the accuracy and efficiency of detecting available spectrum \cite{mishra2006cooperative, wang2010evolutionary, liu2010cognitive}. In cooperative spectrum sensing, the sensing results from the secondary users are shared by all members within the same or neighboring networks. These secondary users then use the collected results to make spectrum access decision collaboratively or individually. If the sensing results are independent from each other, the cooperative spectrum sensing can significantly increase the accuracy of detecting the primary user's activity. Secondary users can learn from others' sensing results to improve their knowledge on the primary user's activity. After the available spectrum is detected, secondary users need to share the spectrum following some predetermined access policy. In general, the more secondary users access the same channel, the less available access time for each of them, i.e., a negative network externality exists in this problem. Therefore, before making decision on spectrum access, a secondary user should estimate both the primary user's activity based on the collected sensing results and the possible number of secondary users accessing the same spectrum.

\subsection{System Model}
\begin{figure}
    \begin{centering}
        \subfigure[Sense]{
        \includegraphics[width=5.5cm]{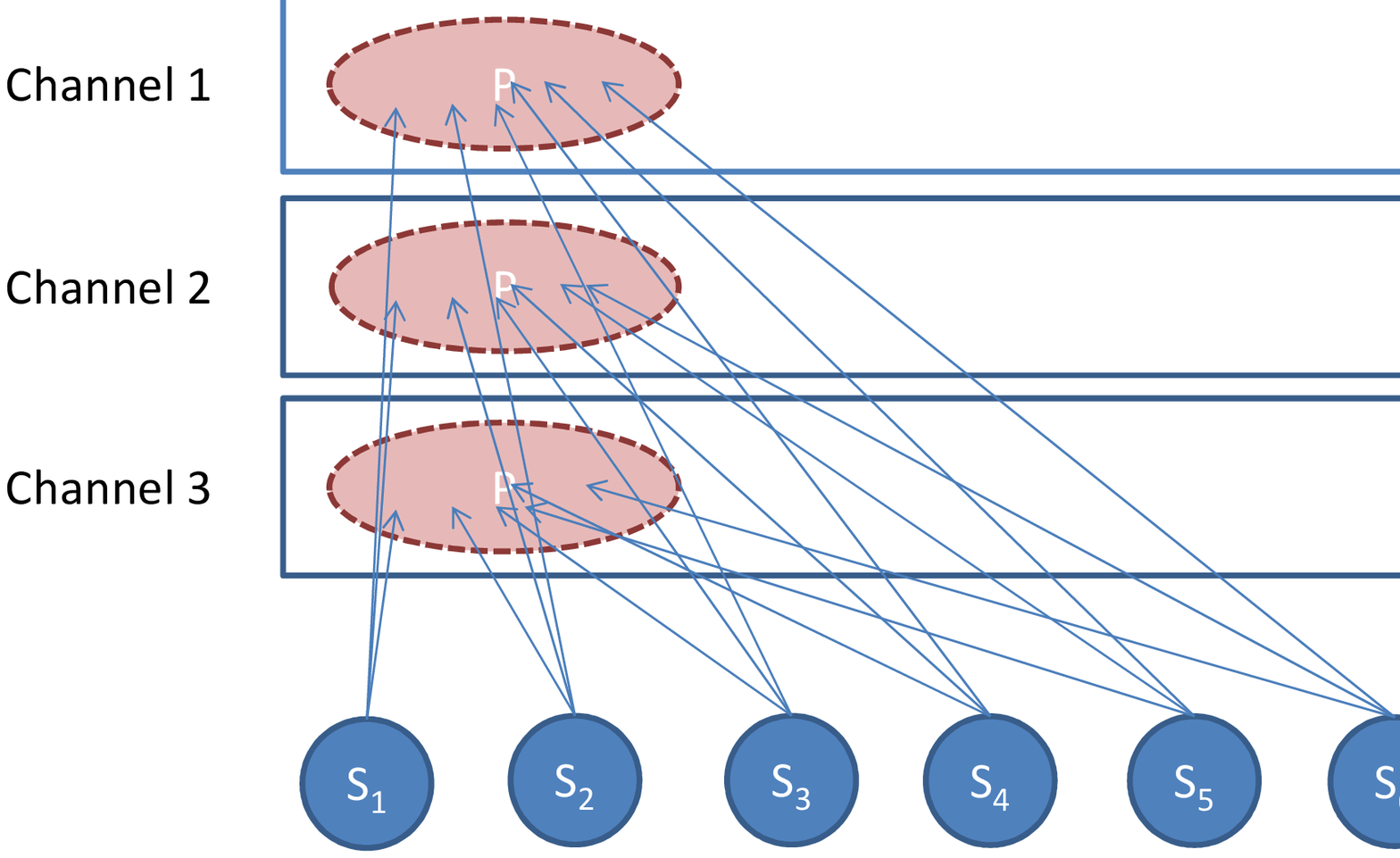}
        \label{fig_cr_1}
      }
      \subfigure[Choose Channels and Broadcast Signals]{
            \includegraphics[width=5.5cm]{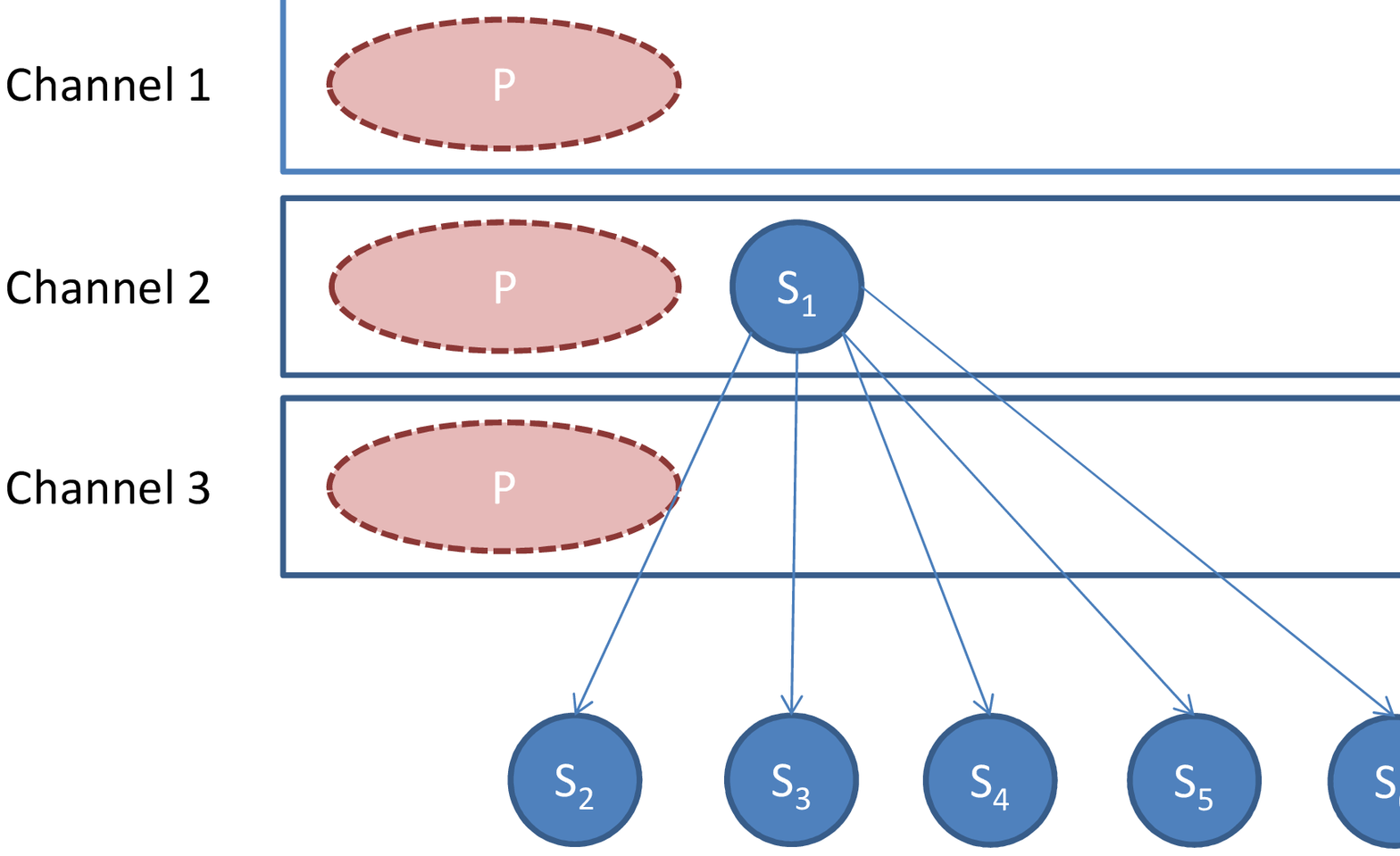}
        \label{fig_cr_2}
      }
      \subfigure[Transmission]{
            \includegraphics[width=5.5cm]{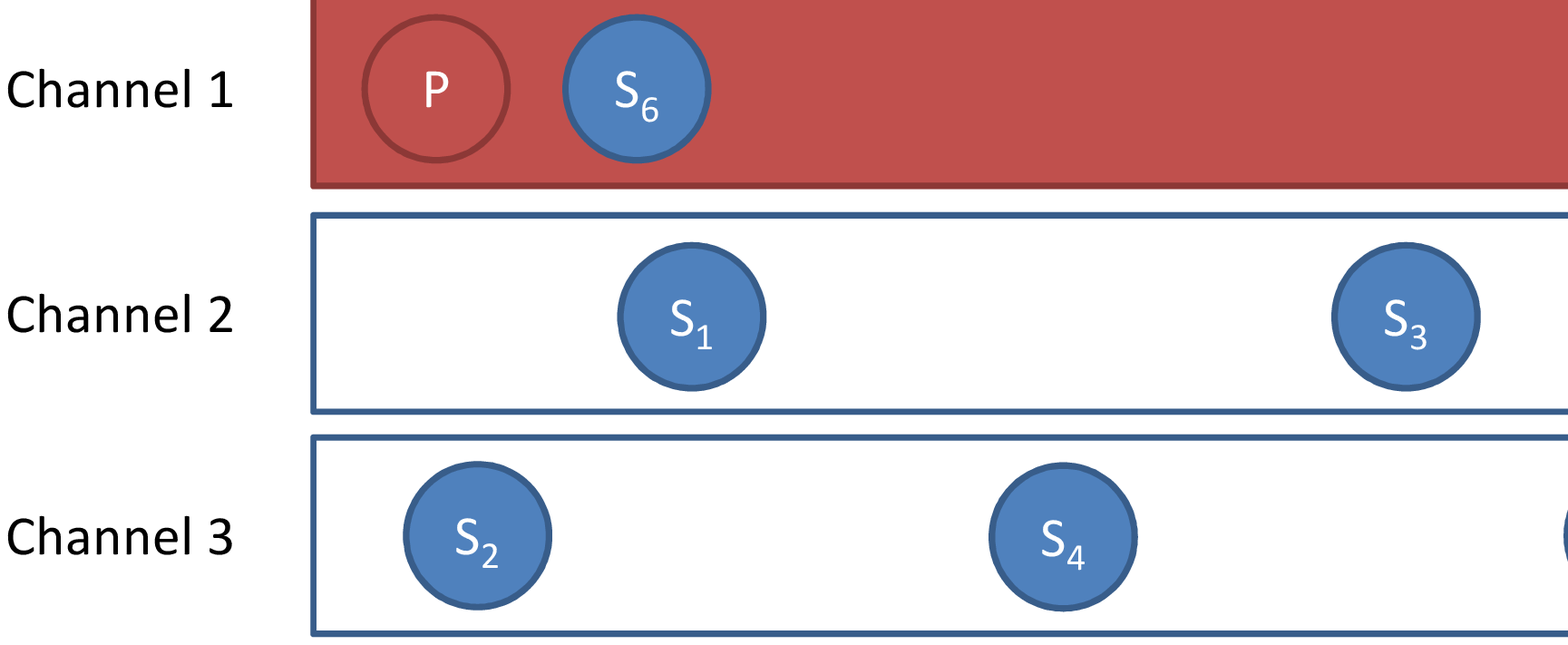}
        \label{fig_cr_3}
      }
    \caption{Sequential Cooperative Spectrum Sensing and Accessing}
    \label{fig_cr}
    \end{centering}
\end{figure}

We consider a cognitive radio system with $K$ channels, $N$ secondary users, and one primary user. We assume that the spectrum access behavior of secondary users is organized by an access point. Suppose that the primary user is always active and transmitting some data on one of the channels. In addition, the primary user's access time is slotted. At each time slot, each channel has equal probability of $1/K$ to be selected by the primary user for transmission. The secondary users' activities are shown in Fig. \ref{fig_cr}. At the beginning of each time slot, secondary users individually perform sensing on all channels $1 \sim K$. Then, they follow a predefined order to sequentially determine which channel they are going to access in this time slot. Without loss of generality, we assume they follow the same order as their indices. When making a decision, a secondary user $i$ reports his decision and the sensing result to the access point. At the same time, all secondary users also receive this report by overhearing. After all secondary users have made their decisions, the access point announces the access policy of each channel: secondary users choosing the same channel equally share the slot time. However, if the channel is occupied by the primary user, their transmission will fail due to the interference from primary user's transmission.

Such a cognitive radio system can be modeled as a Chinese restaurant game. Let $H_{j}$ be the hypothesis that channel $j$ is occupied by the primary user. Then, let the sensing results of secondary user $i \in \{1,2,...,N\}$ on channel $j \in \{1,2,...,K\}$ be $s_{i,j}$. We use a simple binary model on the sensing result in this example, where $s_{i,j}=1$ if the secondary user detected some activity on channel $j$ and $s_{i,j}=0$ if no activity is detected on channel $j$. For secondary user $i$, his own sensing results are denoted as $\mathbf{s_i}=\{s_{i,1},s_{i,2},...,s_{i,K}\}$. In addition, the results he collected from the reports of previous users are denoted as $\mathbf{h_i}=\{\mathbf{s_1},\mathbf{s_2},...\mathbf{s_{i-1}}\}$.

We define the belief of a secondary user $i$ on the occupation of channels as $\mathbf{g_i}=\{g_{i,1},g_{i,2},...,g_{i,K}\}$, where $g_{i,j}=Pr(H_{j}|\mathbf{h_i},\mathbf{s_i})$. Let the probability of false alarm and miss detection of the sensing technique on a single channel as $p_f$ and $p_m$, respectively. The probability of $\mathbf{s_i}$ conditioning on $H_j$ is given by
\begin{equation}
    Pr(\mathbf{s_i}|H_j) = p_m^{1-s_{i,j}}(1-p_m)^{s_{i,j}} \prod_{k \in \{1,2,...,K\} \setminus \{j\}} p_f^{s_{i,k}}(1-p_f)^{1-s_{i,k}}.
\end{equation}

Thus, we have the following belief updating rule
\begin{equation}
    g_{i,j}=\frac{Pr(\mathbf{h_i},\mathbf{s_i}|H_{j})Pr(H_{j})}{\sum_{k=1}^K Pr(\mathbf{h_i},\mathbf{s_i}|H_{k})Pr(H_{k})}=\frac{g_{i-1,j} Pr(\mathbf{s_i}|H_{j})}{\sum_{k=1}^K g_{i-1,k} Pr(\mathbf{s_i}|H_{k})}.
\end{equation}

With this rule, the belief of secondary user $i$ is updated when a new sensing result is reported to the access point. The available access time of a channel $j$ within a slot is its slot time, which is denoted as $T$. However, if the channel occupied by primary user, its access time becomes $0$. Thus, we define the access time of channel $j$ as
\begin{equation}
    R_j(H_k) =\left\{
              \begin{array}{ll}
                0, & \hbox{$j=k$}. \\
                T, & \hbox{otherwise}.
              \end{array}
            \right.
\end{equation}

Then, let $x_i$ be secondary user $i$'s choice on the channels, and $n_j$ be the number of secondary users choosing channel $j$. We define the utility of a secondary user $i$ as
\begin{equation}
    u_i(x_i) = \frac{Q_{x_i}R_{x_i}}{n_{x_i}},
\end{equation}
where $Q_{x_i}$ is the channel quality of channel $x_i$. If the channel has higher quality, the secondary users choosing the channel have higher data rates, and thus higher utility. Then, the best response of secondary user $i$ is as follows,
\begin{equation}
    BE_i(\mathbf{n_i},\mathbf{h_i},\mathbf{s_i}) = \arg \max_x \sum_{k \in \{1,2,...,K\} \setminus \{x\}} g_{i,k} E\left[\frac{Q_x T}{n_x}|\mathbf{n_i},\mathbf{h_i},\mathbf{s_i},H_k \right].
\end{equation}
This best response function can be solved recursively through the recursive equations in (\ref{eqn39}) and (\ref{eqn44}).

\subsection{Simulation Results}

\begin{figure}

    \begin{centering}

        \subfigure[Secondary User 1]{
        \includegraphics[width=7cm]{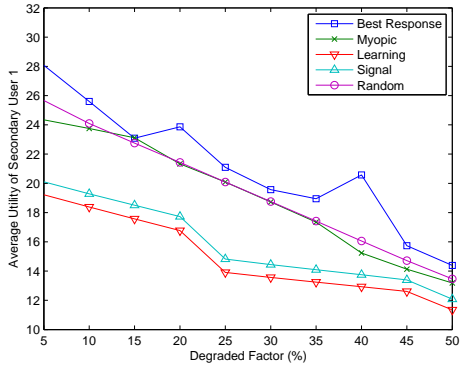}
        \label{fig_sim_cr_1}
      }
      \subfigure[Secondary User 3]{
        \includegraphics[width=7cm]{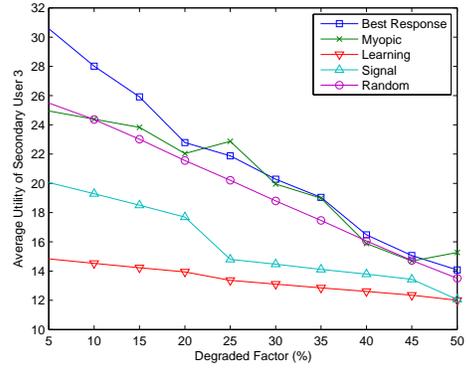}
        \label{fig_sim_cr_3}
      }
      \subfigure[Secondary User 7]{
        \includegraphics[width=7cm]{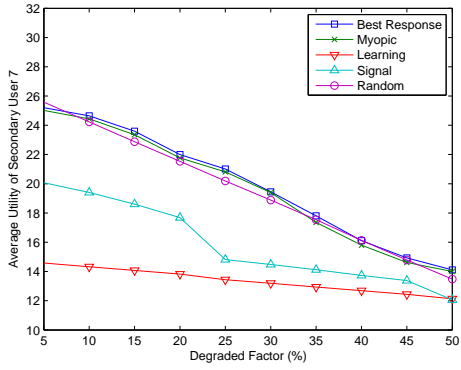}
        \label{fig_sim_cr_7}
      }
      \subfigure[Average Utility of All Secondary Users]{
        \includegraphics[width=7cm]{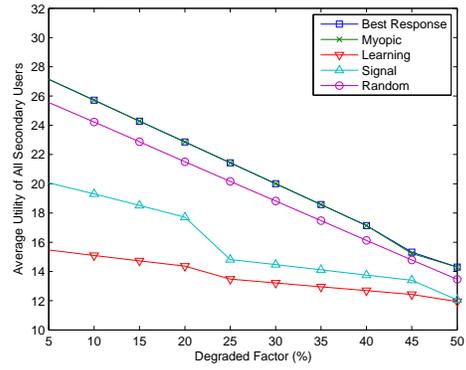}
        \label{fig_sim_cr_avg}
      }
      \subfigure[Secondary Users Interfere the Primary User]{
            \includegraphics[width=7cm]{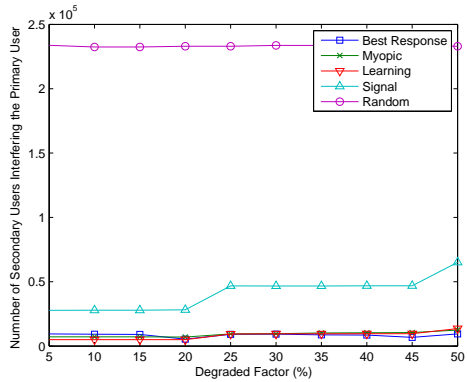}
        \label{fig_sim_cr_miss}
      }

    \caption{Spectrum Accessing in Cognitive Radio Network under Different Schemes}

    \label{fig_sim_cr}

    \end{centering}

\end{figure}

We simulate a cognitive radio network with $3$ channels, $1$ primary user, and $7$ secondary users. When the channel is not occupied by the primary user, the available access time for secondary users in one time slot is $100 ms$. Secondary users sense the primary user's activity in all three channels at the beginning of the time slot. We assume that the primary user has equal probability to occupy one of three channels. Conditioning on the primary user's occupation of the channel, the probabilities of miss detection (if occupied) and false alarm (if not occupied) in sensing one channel are $0.1$. The channel quality factor of channel $1$ is $Q_1=1$, while channel $2$ and $3$ are $1-d$ and $1-2d$. The $d$ is the degraded factor, which is controlled within $[5\%,50\%]$ in the simulations. We compare the proposed best response strategy in (\ref{eqn44}) with other four strategy scheme: random, signal, learning, and myopic strategies. The simulation results are shown in Fig. \ref{fig_sim_cr}.

From Fig. \ref{fig_sim_cr_1}, \ref{fig_sim_cr_3}, and \ref{fig_sim_cr_7}, we can see that secondary users have different utilities under different orders and schemes. For both the myopic and the proposed best response schemes, secondary user $3$ has a larger utility than secondary user $1$ when the degraded factor is low. This is because secondary user $3$ has the advantages in collecting more signals than secondary $1$ to identify the channel occupied by the primary user. Moreover, the loadings of the other two channels are still far from their expected equilibrium loadings since only two secondary users have made choices. Therefore, secondary user $3$ has a larger utility than secondary user $1$.
Nevertheless, when the degraded factor is high, we can see that the utility of secondary user $1$ is larger than that of secondary user $3$.
This is because when the degraded factor increases, the quality difference among channels increases. In such a case, even secondary user $3$ can successfully identify the occupied channel, the channel that offers a higher utility in the equilibrium is usually the one with fewer number of secondary users. The expected number of secondary users accessing such a channel is generally $2$ or even $1$ depending on the degraded factor, and secondary user $3$ can no longer freely choose those channels. For secondary user $7$, who has the best knowledge on the primary user's activity, he usually has no choice since there are six secondary users making decisions before him. Therefore, he has the smallest utility.

Generally, the myopic scheme provides an equal or lower utility than the best response scheme for secondary users making decisions early, such as secondary user $1$, since secondary users in the myopic scheme do not predict the decisions of subsequent users. However, some secondary users eventually benefit from the mistakes made by early secondary users. We can see from Fig. \ref{fig_sim_cr_3} that for the cases that $d=20\%$ and $d=50\%$, customer $3$ has a higher utility under the myopic scheme than under the best response scheme due to the mistakes made by customer $1$ and $2$. We can also see from Fig. \ref{fig_sim_cr_avg} that both best response and myopic schemes provides the same average utilities of all secondary users. In such a case, the utility loss of some secondary users in the myopic scheme will lead to the utility increase of some other secondary users.

For random and signal schemes, there is no difference among the average utilities of secondary user $1$, $3$, and $7$ since secondary users do not learn from other agents' actions and signals under these two schemes. For the learning scheme, we can see that secondary user $1$ has a significantly larger utility than secondary user $3$ and $7$. This is because in the learning scheme, secondary users do not take the negative network externality into account when making decisions on the channel selection. Since secondary users who made decisions later are more likely to identify the primary user's activity, they are more likely to choose the same channels and share with each other, and their utilities are degraded due to the negative network externality.

Let us take a deeper look at the average utility of all secondary users shown in Fig. \ref{fig_sim_cr_avg}. On one hand, we can see that both best response and myopic schemes achieve highest average utilities of all secondary users. The network externality effects in spectrum access force strategic secondary users to access different channels instead of accessing the same high quality channels. On the other hand, learning and signal schemes lead to poor average utilities since they do not consider the network externality in their decision processes. All secondary users tend to access the same available high quality channel, and therefore the spectrum resource in other available channels is wasted. This also explains the phenomenon that learning scheme leads to poorer performance than signal scheme. Under the learning scheme, secondary users are more likely to reach a consensus on the primary user's activity and make the same choice on the channels, which degrades the overall system performance.

Finally, we show the number of secondary users causing interference to the primary user in Fig. \ref{fig_sim_cr_miss}. We can see that those schemes involving learning, which are best response, myopic, and learning schemes, can significantly reduce the interference to the primary user. Secondary users who learn from others' signals efficiently avoid the channel occupied by the primary user. 

\section{Cloud Computing: Cloud Storage Service Selection}\label{sec_app2}
In the second application, we consider the storage service selection in cloud computing. Reliability and availability are the major concerns of potential cloud subscribers in choosing the cloud storage platform. These two features may be enhanced by the cloud computing platform through optimizing configuration, upgrading the software architecture, and hardware infrastructure \cite{Oppenheimer2003wis}. However, reliability and availability are also affected by the number of subscribers. For instance, when overwhelming number of subscribers access the storage service, these subscribers may experience unacceptable waiting time, service blocking, or even data loss in transmission due to network congestion and capacity overloading. The reliability of the platform may be decreased to an unacceptable level even with upgraded infrastructure \cite{Singh2007failover}. In general, for a fixed software architecture and hardware infrastructure, the more subscribers using the same platform, the lower the service quality of the cloud storage platform. Thus, before choosing the cloud service, a potential subscriber should consider both the infrastructure of the platform and the potential growth of the subscription numbers of the platform.

\begin{figure}
    \begin{centering}
    \includegraphics[width=9cm]{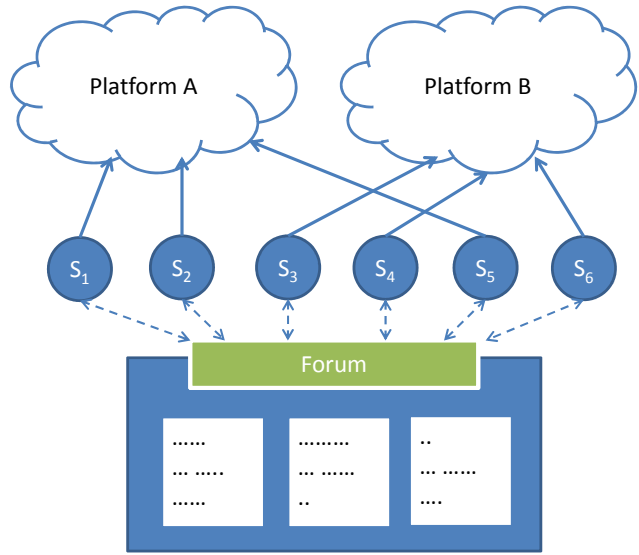}
        \caption{Platform Selection in Cloud Computing}
        \label{fig_cl}
    \end{centering}
\end{figure}

\subsection{System Model}
Let us consider two cloud storage platforms with $N$ potential subscribers. These two platforms, say $A$ and $B$, offer the same storage plan. A storage plan contains two items: maximum storage space and pricing. These plans require long-term contracts, which means that a subscriber cannot change from one platform to another in a short period. Since both platforms provide the same plan, a subscriber may not choose one of them due to the price or the storage size. Suppose that the service qualities of these two platforms are different. One of the platforms has upgraded their infrastructure and can provide higher availability. Let us consider a binary model on the infrastructure, i.e., $H_h$ and $H_l$ are high-reliable infrastructure and low-reliable infrastructure, respectively. When high-reliable infrastructure $H_h$ is used in the platform, the probability that one subscriber causing the service to fatal crash is $p_h$. If the crash happens, the platform becomes unavailable to all subscribers at a certain time. When low-reliable infrastructure $H_l$ is used, the probability of a fatal crash per subscriber is $p_l$, where $p_l > p_h$. We assume that there is no third-party to tell the subscribers what exactly the infrastructure the platforms are using. Thus, both platform may claim that they have upgraded the infrastructure, and the subscribers may not be able to verify it before choosing their services. However, they all know the prior probability $g_0$ that platform $A$ is the one that upgrades the infrastructure. They also have collected some rumors about the true state of each platform, and have formed their own believes on the infrastructure of each platform.

We assume these subscribers make their choices sequentially. When a subscriber $i$ makes his choice, as shown in Fig. \ref{fig_cl}, he announces his decision and posts the rumors he received on a public discussion forum. Thus, all subscribers know his decision and get the rumor he received. We assume the decision process is relatively short comparing with the required long-term contracts for using the storage service. Thus, we ignore the transition state and focus on the service availability of the final state, i.e, the time after all subscribers have made their decisions. The service availability experienced by one subscriber is determined by two factors, which are the number of subscribers at the end of the decision process and the infrastructure used by the platform.

Let the rumor received by subscriber $i$ be a random variable $s_i \in \{A,B\}$ conditioning on the infrastructure used by each platform. When $s_i =A$, the subscriber receives a rumor favoring platform $A$, otherwise the rumor favors platform $B$. Let $H_A$ and $H_B$ be the infrastructure using by platform $A$ and $B$, respectively. The probability distribution $f(s_i|H)$ is defined as follows:
\begin{equation}
    f(s_i=x|H_x) = \left\{
              \begin{array}{ll}
                p, & \hbox{$H_x = H_h$}, \\
                1-p, & \hbox{otherwise},
              \end{array}
       \right.
\end{equation}
where $x \in \{A,B\}$ and $p$ is the conditional probability that the rumor is true. Note that since only one platform upgrades the infrastructure, if $H_A = H_h$, then $H_B=H_l$, and vice versa. The belief of a subscriber $i$ is defined as $g_i=Pr(H_A=H_h|\mathbf{h_i},s_i)$, where $\mathbf{h_i}=\{s_1,s_2,...,s_{i-1}\}$ is the collection of the posted rumors on the public discussion forum. The belief updating rule is given as follows:
\begin{eqnarray}
    \nonumber g_i &=& \frac{Pr(\mathbf{h_i},s_i|H_A=H_h)Pr(H_A=H_h)}{Pr(\mathbf{h_i},s_i|H_A=H_h)Pr(H_A=H_h) + Pr(\mathbf{h_i},s_i|H_A=H_l)Pr(H_A=H_l)} \\
    &=& \frac{g_{i-1}f(s_i|H_A=H_h)}{g_{i-1}f(s_i|H_A=H_h) + (1-g_{i-1})f(s_i|H_A=H_l)}.
\end{eqnarray}

The utility function of a subscriber $i$ choosing the platform $x \in \{A,B\}$ is given by
\begin{equation}
    u_i(x) = (1-p_x)^{n_x},
\end{equation}
where $p_x$ is the probability of service fatal crash caused by one subscriber and $n_x$ is the number of subscribers using platform $x$. The utility function $u_i(x)$ describes the probability that the cloud storage service is available, given the number of subscribers and the reliability of infrastructure.

Finally, let $n_{i,x}$ be the number of subscribers choosing platform $x$ before subscriber $i$, we have $n_{i,A}=i-1-n_{i,B}$. Then, the best choice of a subscriber $i$ is
\begin{equation}
    BE_i(n_{i,A},\mathbf{h_i},s_i) = \left\{
              \begin{array}{ll}
                A, & \hbox{$E[(1-p_A)^{n_A}|n_{i,A},\mathbf{h_i},s_i] > E[(1-p_B)^{n_B}|n_{i,A},\mathbf{h_i},s_i]$}. \\
                B, & \hbox{otherwise}.
              \end{array}
       \right.
\end{equation}

This best response function can be solved recursively through the recursive equations in (\ref{eqn39}) and (\ref{eqn44}).

\subsection{Simulation Results}

\begin{figure}
    \begin{centering}
      \subfigure[Subscriber 1]{
        \includegraphics[width=7cm]{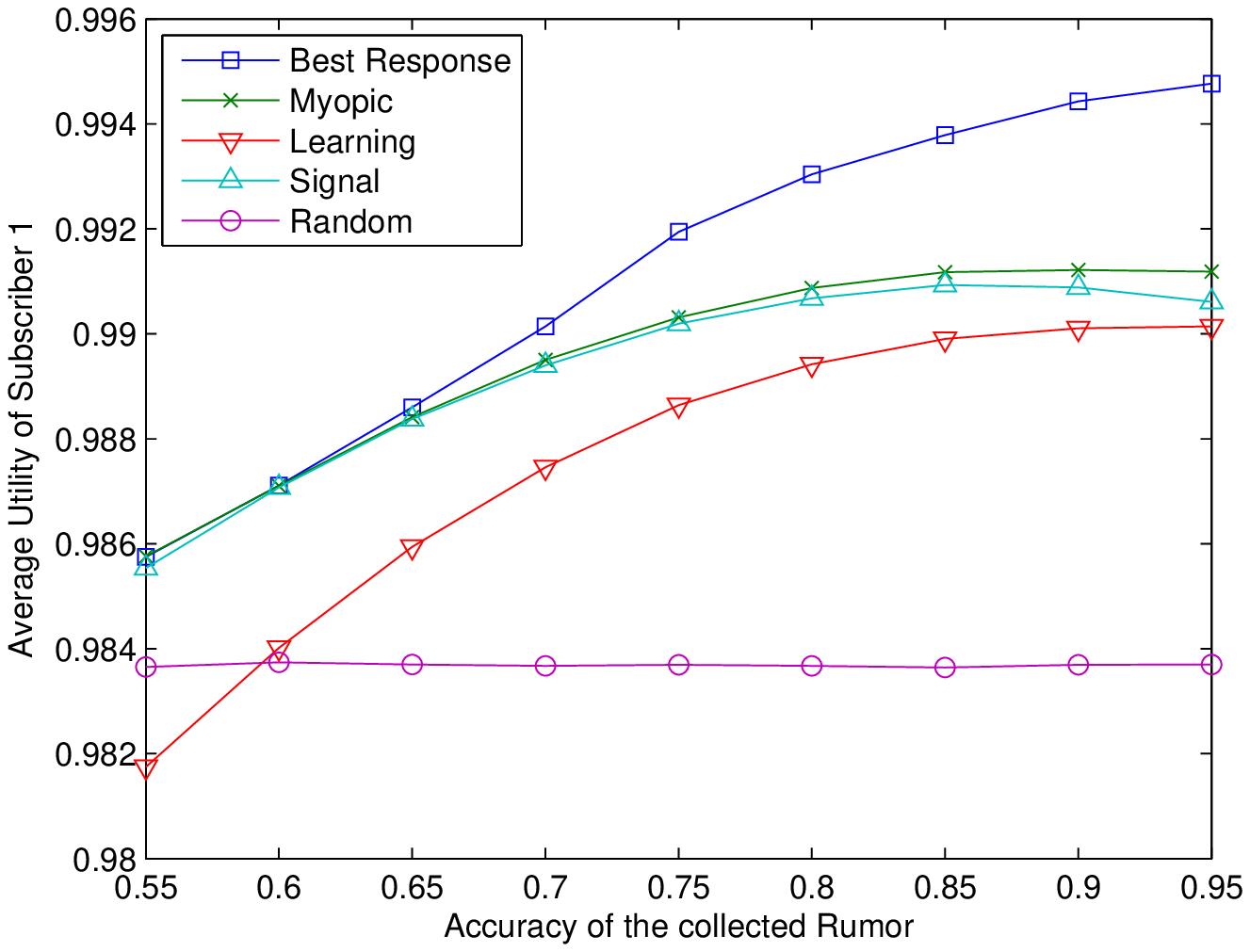}
        \label{fig_sim_cl_q_1}
      }
      \subfigure[Subscriber 10]{
        \includegraphics[width=7cm]{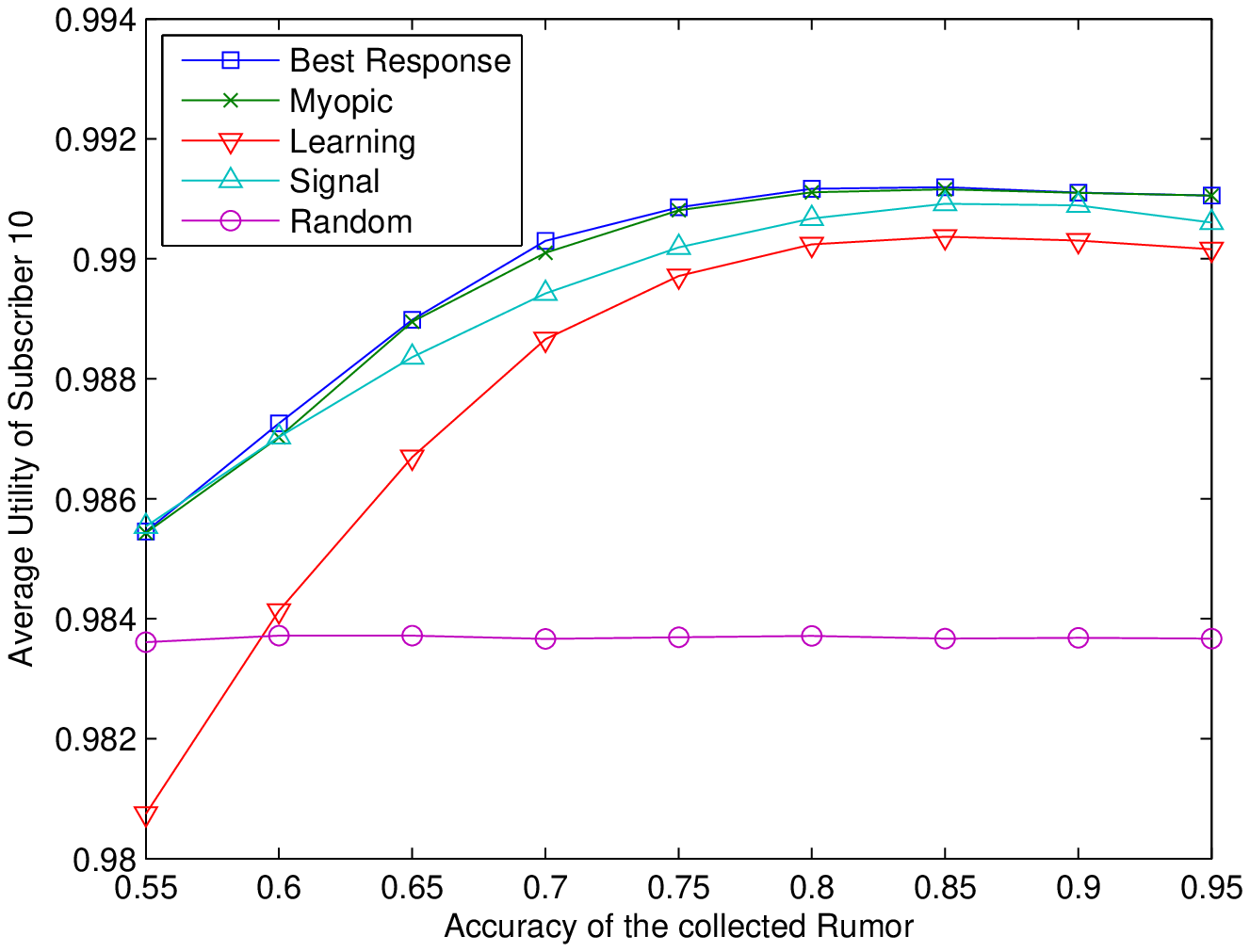}
        \label{fig_sim_cl_q_10}
      }
    \caption{Storage Platform Selection in Cloud Computing vs. Accuracy of Rumors}
    \label{fig_sim_cl_q}
    \end{centering}
\end{figure}

We simulate a system with two storage service platforms and $10$ subscribers. There are two types of infrastructures, which are high-reliable and low-reliable infrastructure. The high-reliable infrastructure $H_h$ offers a service with the probability of failure per user $p_h=0.0001$. For the low-reliable infrastructure, the probability of failure per user is $p_l$. Each subscriber collects a rumor from the Internet, which favors one of these two platforms. In the first simulation, we discuss how subscribers make decisions given different accuracy of the collected rumors. In this simulation, the parameter $p_l$ is set to be 0.0005 and $p$ lies in $[0.55, 0.95]$. The simulation results are shown in Fig. \ref{fig_sim_cl_q}.

From Fig. \ref{fig_sim_cl_q_1}, we can see that the proposed scheme can provide the largest utility for subscriber $1$ and the utility increases as the accuracy of the rumors increases. With the myopic scheme, the utility of subscriber $1$ decreases since the decisions of subsequent subscribers are not taken into account. Nevertheless, by considering the negative network externality effect in the decision process, the myopic scheme performs better than signal and learning schemes. Similar results can be observed for the utility of subscriber $10$, as shown in Fig. \ref{fig_sim_cl_q_10}. The only difference is that the best response scheme provides similar utility to subscriber $10$ compared with the myopic scheme. The reason is that subscriber $10$ is the last subscriber and does not need need to predict any other subscribers' decisions. From Fig. \ref{fig_sim_cl_q_10}, we also observe that the utility of subscriber $10$ with the best response scheme decreases when $p$ is larger than $0.8$. This is because as the accuracy of the rumor increases, the subscribers before subscriber $10$ can better identify the platforms' infrastructures and thus make better decisions. In such a case, subscriber $10$ has a less chance to choose the better platform before it reaches the expected number of subscribers in equilibrium.

\begin{figure}
    \begin{centering}
        \subfigure[Subscriber 1]{
        \includegraphics[width=7cm]{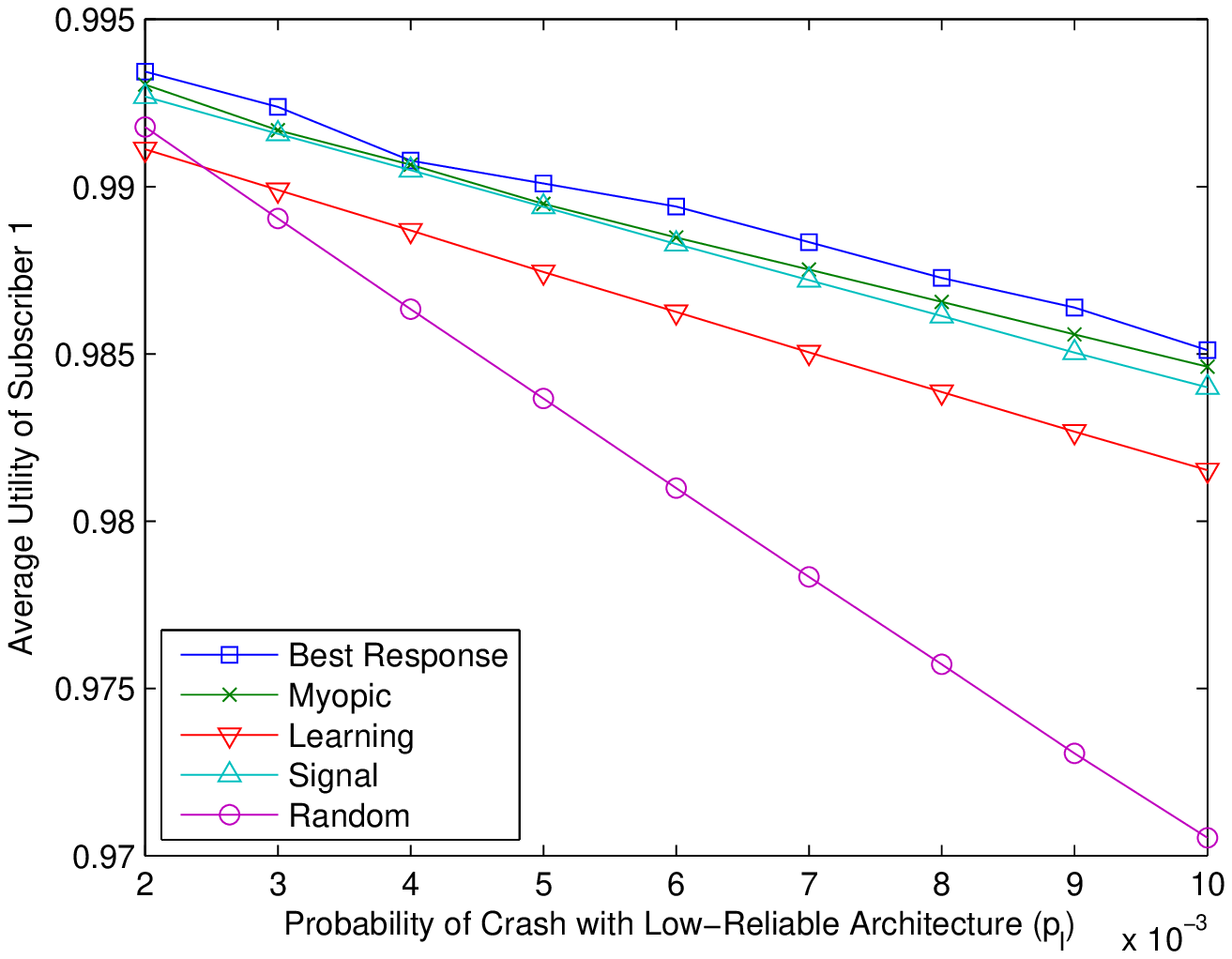}
        \label{fig_sim_cl_1}
      }
      \subfigure[Subscriber 10]{
        \includegraphics[width=7cm]{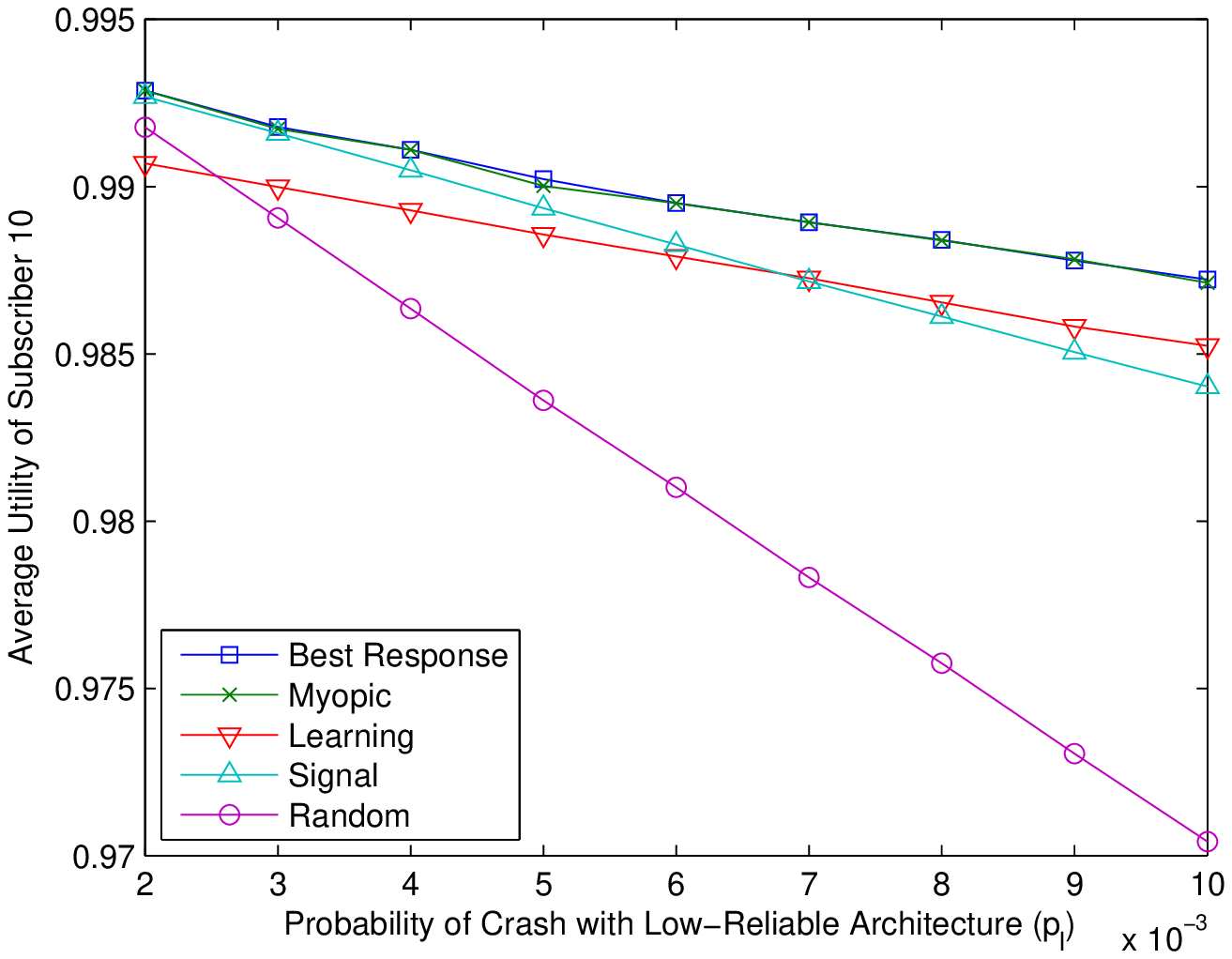}
        \label{fig_sim_cl_10}
      }
      \subfigure[Average Utility of All Subscribers]{
        \includegraphics[width=7cm]{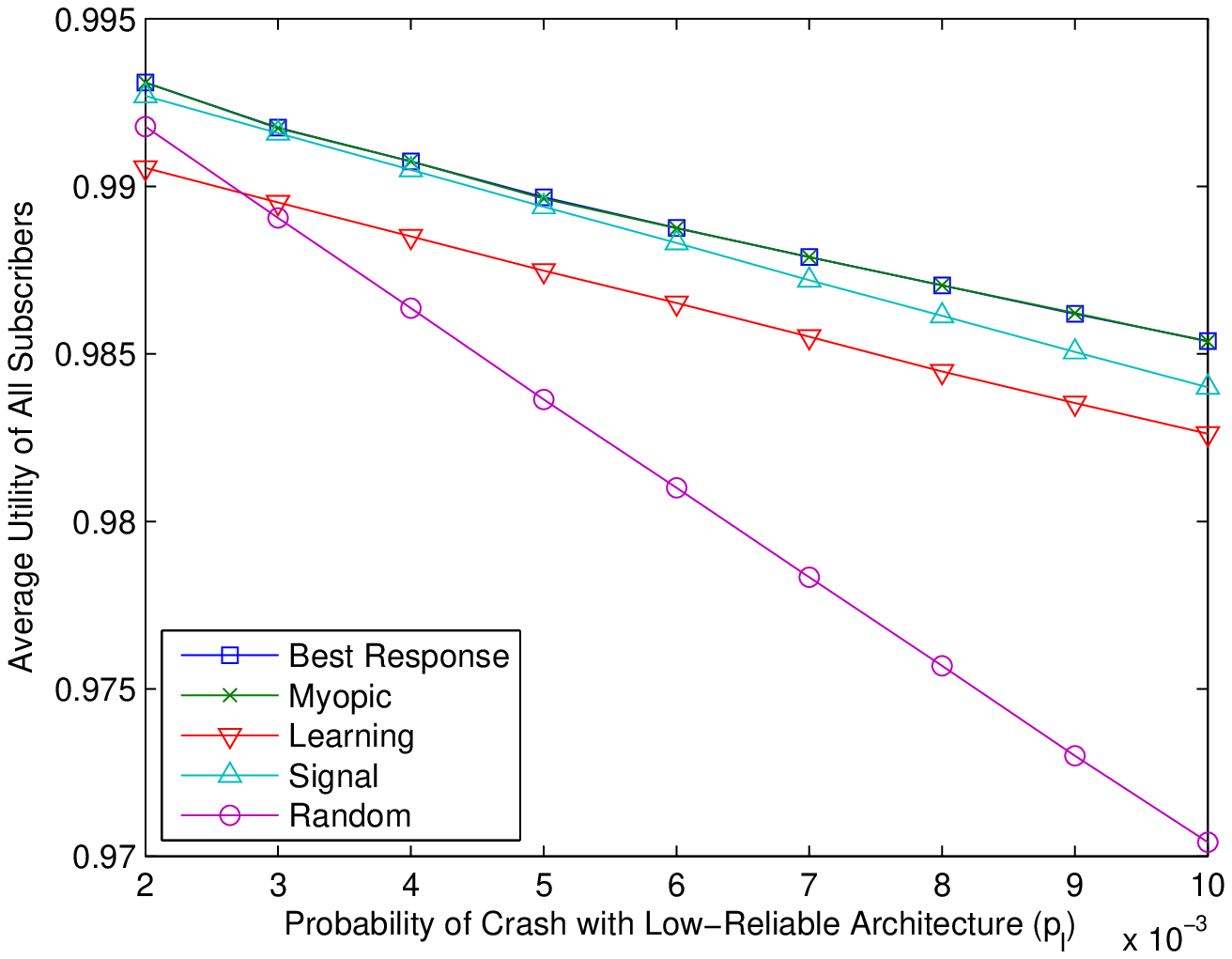}
        \label{fig_sim_cl_avg}
      }
      \subfigure[Reliability under High- and Low-reliable Platforms]{
        \includegraphics[width=7cm]{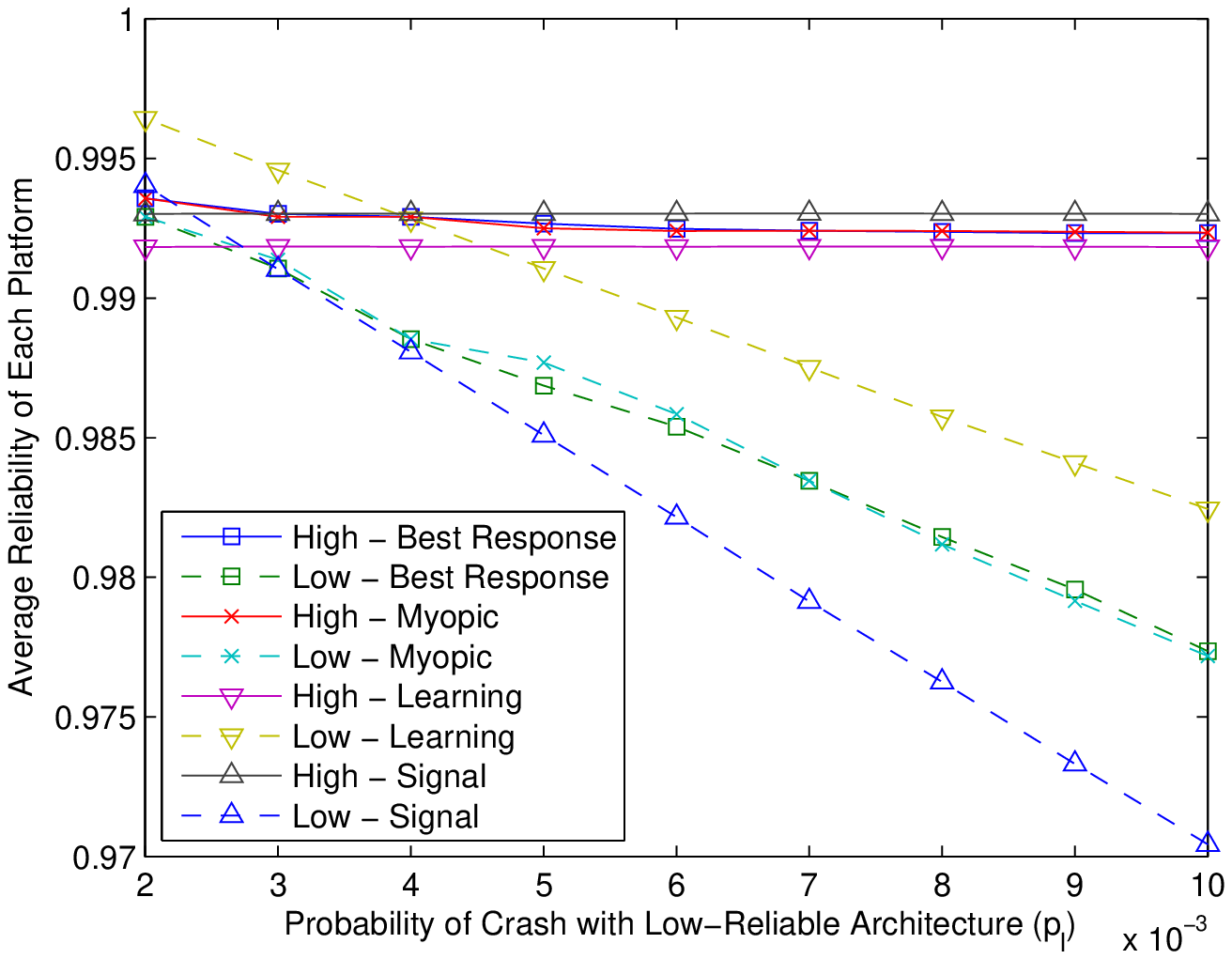}
        \label{fig_sim_cl_p}
      }
    \caption{Storage Platform Selection in Cloud Computing vs. Reliability of Platforms}
    \label{fig_sim_cl}
    \end{centering}

\end{figure}

Then, we simulate with $p_l \in [0.0002,0.001]$ and $p = 0.7$ to see how subscribers make choices given different reliability of low-reliable platform. The simulation results are shown in Fig. \ref{fig_sim_cl}. From Fig. \ref{fig_sim_cl_1} and \ref{fig_sim_cl_10}, we observe that best response scheme gives both subscriber $1$ and subscriber $10$ the largest utilities among all schemes. Moreover, subscriber $1$ has a larger utility than subscriber $10$ when $p_l$ is low, but has a smaller utility when $p_l$ is high. This is because when $p_l$ is low and close to $p_h$, the network externality effect dominates the reliability of the platform. Subscriber $1$ has the advantage to choose the one offering the higher reliability in equilibrium before it is over-crowded. However, when $p_l$ is high, the difference in infrastructure dominates the reliability of the platform. Subscribers would prefer the platform with the high-reliable infrastructure even other subscribers already choose the same platform. In such a case, subscriber $10$ has the advantage to identify the platform with the high-reliable infrastructure with the collected rumors.

The myopic scheme offers a lower utility for subscriber $1$ since it lacks the prediction on the decisions of subsequent subscribers. For subscriber $10$, however, there is almost no difference between these two schemes. This is because subscriber $10$, as the last subscriber, does not have to predict the decisions of other subscribers. For signal and learning schemes, subscribers have lower utilities since they do not take the network externality into account. In most cases, the signal scheme provides a larger utility to subscribers than the learning scheme since subscribers achieve some kind of load balance between these two platforms according to the signal distribution $f(s|\theta)$. Note that when $p_l$ is high, the utility of subscriber $10$ under the learning scheme is larger than that under the signal scheme. This is because when $p_l$ is high, the platform with high-reliable infrastructure provides higher reliability even it serves more subscribers. In such a case, a subscriber should try to identify and choose the high-reliable platform, which is exactly what the learning strategy does. The average utility of all subscribers under different schemes are shown in Fig. \ref{fig_sim_cl_avg}. We can see that both best response and myopic schemes provide the highest average utility and the signal scheme performs better than the learning scheme due to the load-balancing from the signal distribution $f(s|\theta)$.

In Fig. \ref{fig_sim_cl_p}, we evaluate the average reliability of each platform under different schemes. Note that in Fig. \ref{fig_sim_cl_p} if the platform is not chosen by any subscriber in the simulation, we assume the reliability of that platform is $1$. For the platform with the high-reliable infrastructure, we can see that it has the highest reliability under the signal scheme and the lowest reliability under the learning scheme. On the contrary, the low-reliable infrastructure has the highest reliability under the learning scheme and the lowest reliability under the signal scheme. We can see that both best response and myopic schemes provides a better load balance between these two platforms. The reliability of the low-reliable platform is significantly improved without much loss on the high-reliable platform.

\section{Online Social Networking: Deal Selection on Groupon}\label{sec_app3}
\begin{figure}
    \begin{centering}
    \includegraphics[width=9cm]{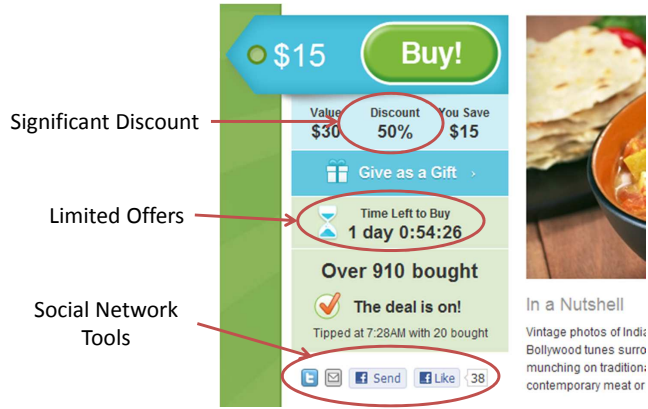}
        \caption{Deals on Groupon}
        \label{fig_gr}
    \end{centering}
\end{figure}

Finally, we consider the effect of negative network externality on online social networking such as Groupon. Recently, a new social network called Groupon shows a new possibility of e-commerce business model \cite{dholakia2010effective}. As shown in Fig. \ref{fig_gr}, it offers small businesses, especially local restaurants, a platform to promote their products with significant discounted deals. These deals are mostly effective only in a limited time or even in a limited amount for advertising purpose. Customers who purchase deals on Groupon may also promote the deal to other social networks like Facebook or Twitter through the built-in tools provided by Groupon. However, it has been observed that when one product or service has been successfully promoted through deals, it is likely to receive negative responses and low reputation due to degraded qualities of services or over-expectation on the products \cite{byers2011daily}. For example, a local restaurant may provide a $50\%$-off deal for advertising purpose. However, this deal may be sold in thousands, which means that the restaurant needs to serve a huge number of customers in months with little profit. In such a case, the quality of meal and service will be degraded, otherwise the restaurant may not be able to survive \cite{tomoko2011groupon}. Thus, a negative network externality exists in this problem. The customer should take into account the possibility of degraded service quality when choosing the deals on Groupon.

\subsection{System Model}
Let us consider $K$ deal sites, each of them offering a deal from a restaurant. These restaurants are of the same kind and in the same area, which means that they share the same group of potential customers. We assume there are $N$ potential customers for these restaurants. The deals these restaurants provide have the similar time limitation, i.e., in general one customer will purchase only one of them. The price of the deal provided by restaurant $j$, $c_j$, is known by all customers. On the contrary, the quality of meals in the restaurant $j$, $Q_j$, is unknown since these customers have not visited the restaurant. However, they may collect some reviews on the Internet to estimate $Q_j$. We model $Q_j$ with a binary model, i.e., $Q_j \in \{Q_h,Q_l\}$ where $Q_h$ and $Q_l$ denote high and low quality, respectively. We also assume the quality of restaurants are uncorrelated. The $N$ customers purchase deals sequentially. After one customer purchases a deal, his purchase along with the review he found will be posted on some public social networks that can be seen by all customers. Finally, after all purchases have been made, each customer visits the restaurant according to the deal he purchased for a meal.

The review customer $i$ found on one restaurant $j$, denoted as $s_{i,j}$, may be positive $s_p$ or negative $s_n$ in probability. Conditioning on the quality of the restaurant $Q_j$, the probability distribution is described as follows:
\begin{equation}
    f(s_{i,j}|Q_j) = \left\{
            \begin{array}{ll}
                p, & \hbox{$s_{i,j}=s_p, Q_j = Q_h$ or $s_{i,j}=s_n, Q_j = Q_l$}, \\
                1-p, & \hbox{otherwise},
              \end{array}
       \right.
\end{equation}
where the $p$ represents the quality of reviews. We denote $\mathbf{s_i}=\{s_{i,1},s_{i,2},...,s_{i,K}\}$ as the reviews customer $i$ collected for all restaurants and denote $\mathbf{h_i}=\{\mathbf{s_1},\mathbf{s_2},...,\mathbf{s_{i-1}}\}$ as the reviews shared by previous customers $1 \sim i-1$. Then, the belief of a customer $i$ on restaurant $j$'s quality is given as follows:
\begin{eqnarray}
    \nonumber g_{i,j} &=& \frac{Pr(\mathbf{h_i},\mathbf{s_i}|Q_j=Q_h)Pr(Q_j=Q_h)}{Pr(\mathbf{h_i},\mathbf{s_i}|Q_j=Q_h)Pr(Q_j=Q_h)+Pr(\mathbf{h_i},\mathbf{s_i}|Q_j=Q_l)Pr(Q_j=Q_l)}\\
    & =& \frac{g_{i-1,j} f(s_{i,j}|Q_j=Q_h)}{g_{i-1,j} f(s_{i,j}|Q_j=Q_h)+(1-g_{i-1,j}) f(s_{i,j}|Q_j=Q_l)}.
\end{eqnarray}
Note that since the quality of restaurants are uncorrelated, the belief on one restaurant is only determined by the reviews related to it.

Finally, let $n_j$ be the number of customers purchasing restaurant $j$'s deal, $x_i$ be the choice of customer $i$, then the utility function of customer $i$ can be defined as
\begin{equation}
    u_i(Q_{x_i},n_{x_i},c_{x_i}) = R(Q_{x_i})- d n_{x_i}- c_{x_i},
\end{equation}
where $R(Q)$ is the value function of a customer on the quality of the meal and $d$ is the crowd discount factor. Here, we assume the degradation of restaurant quality is linear to the number of customers. According to the utility function, we have
\begin{equation}\label{eq_linear_util}
    E[u_i(Q_j,n_j,c_j)|\mathbf{n_i},\mathbf{h_i},\mathbf{s_i},x_i=j] = E[R(Q_j)|\mathbf{h_i},\mathbf{s_i}] - d E[n_j|\mathbf{n_i},\mathbf{h_i},\mathbf{s_i},x_i=j] - c_j.
\end{equation}

\subsection{Simplification with Linear Utility Function}
With the linear utility function in (\ref{eq_linear_util}), we can significantly simplify the original recursive solution. Due to the linearity of the utility function, the expected utility is determined by the expected value of $R(Q_j)$ and the expected number of customers choosing $j$. The expected value of $R(Q_j)$ can be derived directly through current belief of the customer $i$ as
\begin{equation}\label{eq_linear_r}
    E[R(Q_j)|\mathbf{h_i},\mathbf{s_i}] = g_{i,j} R(Q_h) + (1-g_{i,j}) R(Q_l).
\end{equation}

For the expected number of customers choosing $j$, we first define $m_{i,j}$ as the number of customers choosing $j$ after customer $i$ (including customer $i$ himself). Then, we have
\begin{equation}\label{eq_linear_n}
    E[n_j|\mathbf{n_i},\mathbf{h_i},\mathbf{s_i}] = n_{i,j} + E[m_{i,j}|\mathbf{n_i},\mathbf{h_i},\mathbf{s_i}].
\end{equation}

According to (\ref{eq_linear_util}), (\ref{eq_linear_r}), and (\ref{eq_linear_n}), the best response function of customer $i$ is given by
\begin{equation}\label{eq_linear_be}
    BE_i(\mathbf{n_i},\mathbf{h_i},\mathbf{s_i}) = \arg \max_j \{g_{i,j} R(Q_h) + (1-g_{i,j}) R(Q_l) - d n_{i,j} - c_j - d E[m_{i,j}|\mathbf{n_i},\mathbf{h_i},\mathbf{s_i},x_i=j]\}.
\end{equation}

In (\ref{eq_linear_be}), the only unknown term is $E[m_{i,j}|n_A^i,\mathbf{h_i},\mathbf{s_i},x_i=j]$, which can be calculated recursively. Suppose that we have the best response function of customer $i+1$. The best response of customer $i+1$ is $x_{i+1} = j \text{ if and only if }\mathbf{s_{i+1}} \in  S_{i+1,j}(\mathbf{n_{i+1}},\mathbf{h_{i+1}})$.

Conditioning on the current belief $\mathbf{g_i}$, the probability distribution of reviews $f(\mathbf{s}|\mathbf{g_i})$ is given as
\begin{equation}
    f(\mathbf{s}|\mathbf{g_i}) = \prod_{j=1}^K \frac{g_{i,j}f(s_j|Q_j = Q_h)}{g_{i,j}f(s_j|Q_j = Q_h) + (1-g_{i,j})f(s_j|Q_j = Q_l)}
\end{equation}

Then, the recursive form to compute $E[m_{i,j}|\mathbf{n_i},\mathbf{h_i},\mathbf{s_i},x_i]$ is given as follows:
\begin{eqnarray}\label{eqn52}
\!\!\!\!\!\!\!\!&&\!\!\!\!\!\!\!\!E[m_{i,j}|\mathbf{n_i},\mathbf{h_i},\mathbf{s_i},x_i]=\left\{
                                        \begin{array}{ll}
                                          1+E[m_{i+1,j}|\mathbf{n_{i}},\mathbf{h_i},\mathbf{s_i},x_i], & \hbox{$x_i=j$,} \\
                                          E[m_{i+1,j}|\mathbf{n_{i}},\mathbf{h_i},\mathbf{s_i},x_i], & \hbox{$x_i \neq j$,}
                                        \end{array}
                                      \right.\nonumber \\
                                      \!\!\!\!\!\!\!\!&&\!\!\!\!\!\!\!\!=\left\{
                                        \begin{array}{ll}
                                          1 + \sum_{u \in K} \int_{s \in S_{i+1,u}(\mathbf{n_{i+1}},\mathbf{h_{i+1}})} {E[m_{i+1,j}|\mathbf{n_{i+1}},\mathbf{h_{i+1}},\mathbf{s},x_{i+1}=u]f(\mathbf{s}|\mathbf{g_i})ds}, & \hbox{$x_i=j$,} \\
                                          \sum_{u \in K} \int_{s \in S_{i+1,u}(\mathbf{n_{i+1}},\mathbf{h_{i+1}})} {E[m_{i+1,j}|\mathbf{n_{i+1}},\mathbf{h_{i+1}},\mathbf{s},x_{i+1}=u]f(\mathbf{s}|\mathbf{g_i})ds}, & \hbox{$x_i \neq j$,}
                                        \end{array}
                                      \right.
\end{eqnarray}
with $\mathbf{h_{i+1}}$ and $\mathbf{n_{i+1}}$ being defined in (\ref{eq_h_plus}) and (\ref{eq_n_plus}).

\subsection{Simulation Results}
\begin{figure}
    \begin{centering}
        \subfigure[Customer 1]{
        \includegraphics[width=7cm]{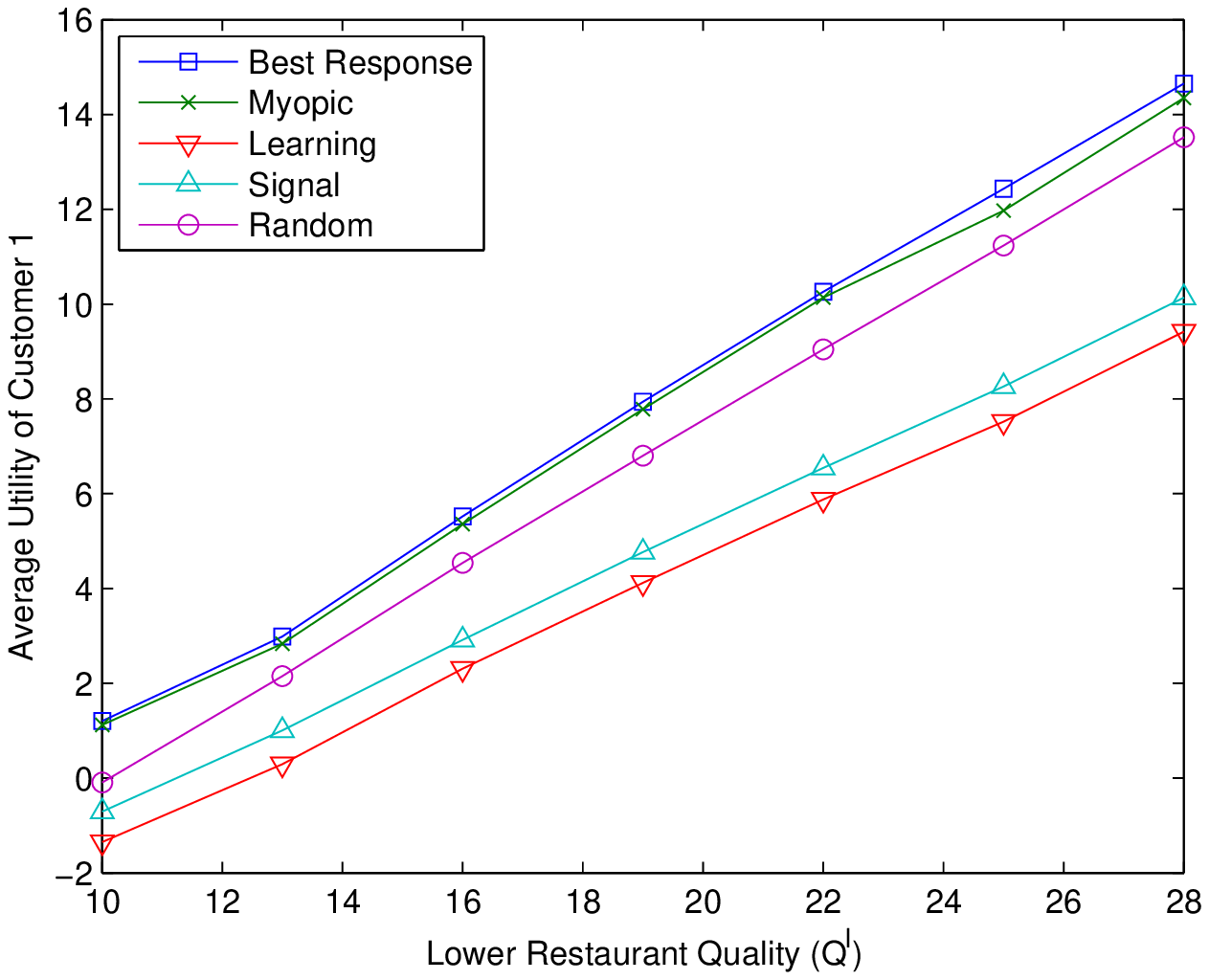}
        \label{fig_sim_gr_1}
      }
      \subfigure[Customer 9]{
        \includegraphics[width=7cm]{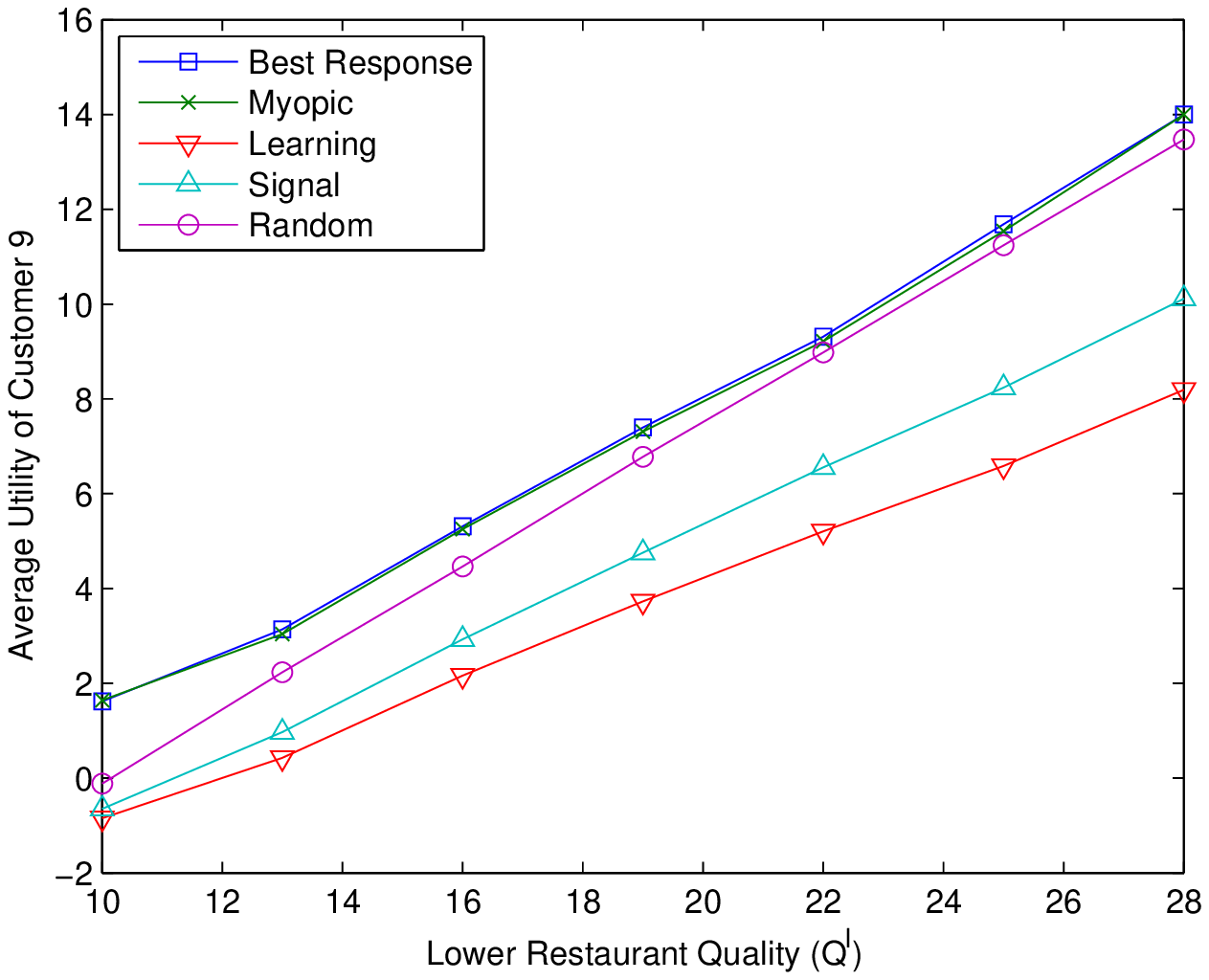}
        \label{fig_sim_gr_9}
      }
      \subfigure[Average Utility of All Customers]{
        \includegraphics[width=7cm]{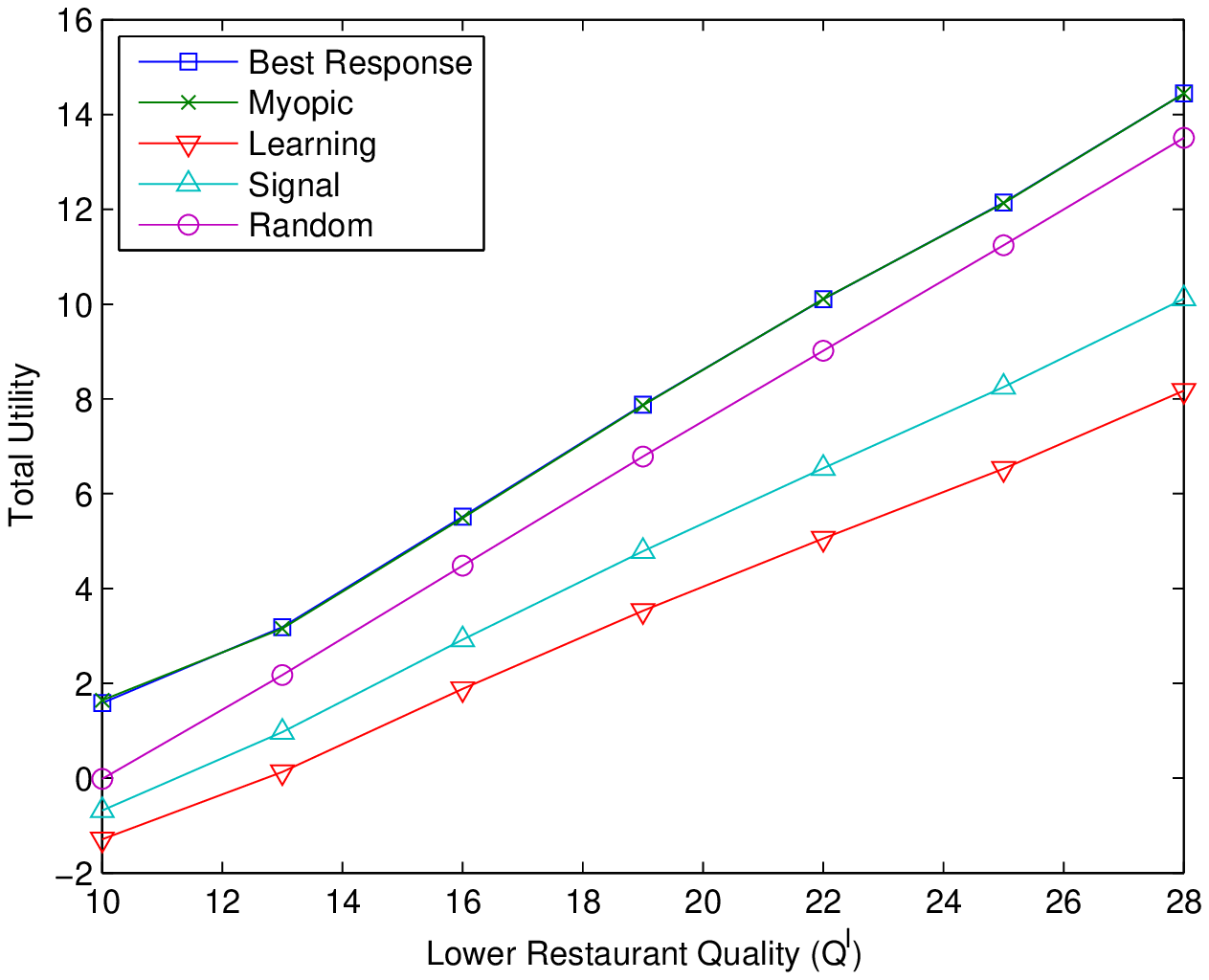}
        \label{fig_sim_gr_avg}
      }
    \caption{Deal Selection in Groupon under Different Schemes}
    \label{fig_sim_gr}
    \end{centering}

\end{figure}

We simulate a deal-offering website with two deals from two different restaurants and $9$ customers. These two deals offer similar meals and have the same price of $5$. We assume that there are two types of restaurant, which are the high quality restaurant with $Q_h=30$ and the low quality restaurant with $Q_l \in [10,28]$. Let the crowd discounting factor $d$ in the utility function be $2$. The probability that one restaurant's quality is high or low are equal. Moreover, a restaurant's quality is independent from the other restaurant's quality. Conditioning on the restaurant's quality, a customer receives a positive review on the restaurant with a probability of $0.7$ if the restaurant's quality is high. Similarly , if the restaurant's quality is low, the probability that a customer receives a negative review on the restaurant is also $0.7$. Each customer receives a review at the beginning of the simulation. Then, they choose the deals and reveal their collected reviews to other customers sequentially. We compare the best response strategy in (\ref{eqn44}) with other four strategies: random, signal, learning, and myopic strategies. The simulation results are shown in Fig. \ref{fig_sim_gr}.

As shown in Fig. \ref{fig_sim_gr_1} and \ref{fig_sim_gr_9}, we can see that customers with different orders have different average utilities except the random scheme. For the best response scheme, when $Q_l$ is low, customer $1$ has the smaller average utility than customer $9$. This is because the restaurant quality difference between high and low is more significant when $Q_l$ is low, and the major factor that determines the utility in this case is the quality of the restaurant. Since customer $9$ has the advantage to collect more reviews and know the quality of restaurants with a stronger belief, he can choose more wisely on the deals and thus larger utility. On the contrary, when $Q_l$ is high, the difference between the restaurant quality is small, and the network externality becomes the major factor in determining the utility of the customers. In such a case, early customer, such as customer $1$, have the advantage to choose the restaurant that has a lower number of customers at the equilibrium.

The myopic scheme provides a smaller utility for customer $1$ than the best response scheme since the expected decisions of subsequent customers are not taken into account under this scheme. For customer $9$, who is the last customer in the simulation, the utility difference becomes indistinguishable between myopic and best response scheme. We can also see that customers have the lowest utility under the learning scheme. This is because all customers are likely to choose the same deal without considering the network externality effect under the learning scheme, which significantly reduces the quality of the meals and thus smaller utilities. This in some sense also reflects the phenomenon we previously mentioned that the combination of traditional social learning and deal promotion on Groupon may eventually reduce the quality of the products if the customers make decisions from learning without considering the network externality effect \cite{byers2011daily, tomoko2011groupon}. Customers with the signal scheme have higher utilities than with the learning scheme since some customers may receive different signals and thus make different decisions. The random scheme performs better than signal and learning schemes in the simulations since it equally separates the crowd in probability and therefore prevents the severe quality degradation from the negative network externality. However, it does not perform better than myopic and best response scheme since both of these schemes jointly consider the network externality and the social learning.

\subsection{Deal Pricing}
Finally, we would like to study how a new restaurant should price their deals to promote himself and maximize the revenue given other existing restaurants through the Chinese restaurant game. Let us consider a new restaurant who tries to enter the market. The owner would like to promote his restaurant through putting a new deal on the Groupon with a discounted price. The restaurant's quality is not known by the customers before opening, but the restaurant may do some advertisements or invite reviewers to give reviews on the restaurant's quality. These become signals to potential customers about the quality of the restaurant. Note that both methods can be controlled by the restaurant, i.e, the signal quality can be controlled by the new restaurant. Therefore, the owner need to not only determine the deal price but also control the signal quality. Depending on the true quality of this new restaurant, the optimal deal price and the signal quality can be determined through the Chinese restaurant game.

\begin{figure}
    \begin{centering}
        \subfigure[Customers choosing New Restaurant with Low Quality]{
        \includegraphics[width=7cm]{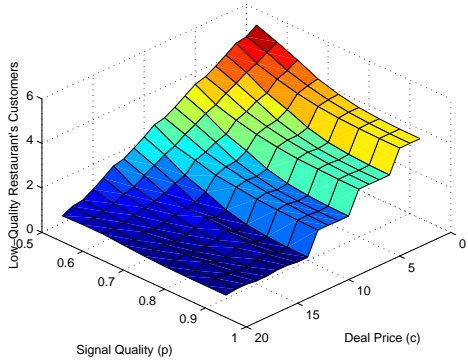}
        \label{fig_sim_gr_price_n_l}
      }
      \subfigure[Customers choosing New Restaurant with High Quality]{
        \includegraphics[width=7cm]{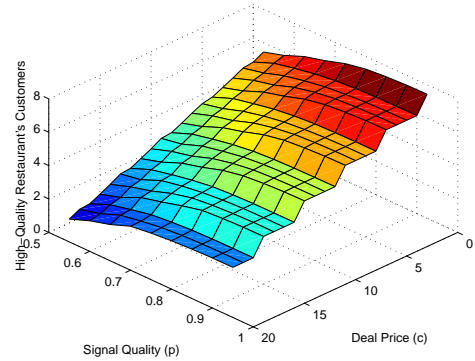}
        \label{fig_sim_gr_price_n_h}
      }
      \subfigure[Revenue of New Restaurant with Low Quality]{
        \includegraphics[width=7cm]{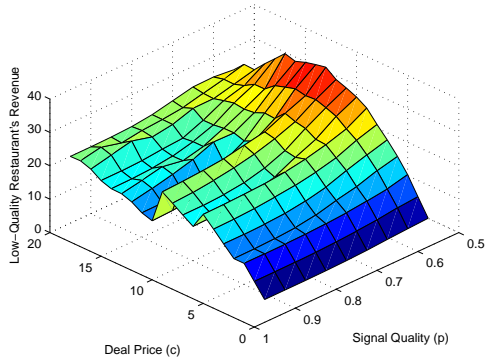}
        \label{fig_sim_gr_price_l}
      }
      \subfigure[Revenue of New Restaurant with High Quality]{
        \includegraphics[width=7cm]{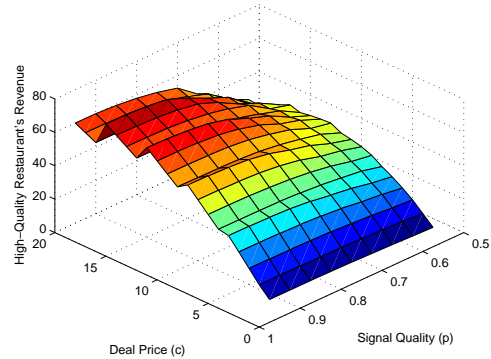}
        \label{fig_sim_gr_price_h}
      }
    \caption{Influence of Deal Price and Signal Quality on the Number of Customers and Revenue of New Restaurant }
    \label{fig_sim_gr_price}
    \end{centering}
\end{figure}

Here, we provide a numerical analysis on how a restaurant should determine the deal price and the signal quality. We consider a deal-offering website with two deals, one from an existing restaurants $A$, and one from a new restaurant $B$. We assume that there are two types of restaurant, which are the high quality restaurant with $Q_h=25$ and the low quality restaurant with $Q_l=10$.  The quality of restaurant $A$ is $Q_h$, which is already known by all customers. Nevertheless, the new restaurant's quality is unknown to all customers. The probabilities that this new restaurant's quality is high or low are equal. Conditioning on the new restaurant's quality, a customer receives a positive review on the restaurant with a probability of $p$ if the restaurant's quality is high. Similarly, if the restaurant's quality is low, the probability that a customer receives a negative review on the restaurant is also $p$. Note that $p$ is controlled by the new restaurant as we discussed before. Suppose that the price of the deal offered by the existing restaurant is $10$. For the deal offered by the new restaurant, its price is denoted as $c$. All other settings are the same as previous simulations. We examine the expected revenue of the new restaurant through simulations. Let $p \in [0, 0.5]$ and $c \in [1, 20]$, the expected number of customers and revenue given the quality of the new restaurant are shown in Fig. \ref{fig_sim_gr_price}.

From Fig. \ref{fig_sim_gr_price_n_l} and \ref{fig_sim_gr_price_n_h}, we can see that the number of customers choosing the new restaurant always increases as the deal price is lower, which is intuitive since a lower price means a higher utility to all customers regardless the restaurant's quality. The signal quality, on the other hand, has a contradictory effect on the restaurant. We can see that when the signal quality increases, the number of customers choosing the new restaurant increases if and only if it is high-quality. Otherwise, the number of customers decreases when the signal quality increases. This is because when the signal quality is high, customers are more likely to identify the true quality of the restaurants. If the restaurant is high-quality, they are more willing to choose the new deal. However, if the restaurant is low-quality, they would rather avoid it. This suggests that if the restaurant is high-quality, the owner should try to increase the signal quality as much as possible by advertising or providing reviews. On the other hand, if the restaurant is low-quality, the owner may provide no information about the quality of this new restaurant.

Considering the revenue of the restaurant, as shown in Fig. \ref{fig_sim_gr_price_l} and \ref{fig_sim_gr_price_h}, the optimal deal prices are different for low- and high-quality restaurants. From the simulations, we can see that the optimal price for the high-quality restaurant is higher than the low-quality restaurant. This is because even the deal price is higher, customers who identify the true quality of the restaurant are still willing to choose this restaurant for higher quality of meals. We also observe that a higher signal quality increases the revenue of the high-quality restaurant but decreases the revenue of the low-quality restaurant. Therefore, a high-quality new restaurant should try to improve the signal quality in order to make potential customers identify its true quality. On the contrary, a low-quality new restaurant may try to hide his information and provide a lower deal price in order to attract more customers and thus lead to a higher revenue.

\section{Conclusion}\label{sec_con}

In Part I of this two-part paper \cite{wang2011crgpart1}, we had proposed a new game, called Chinese restaurant game, to analyze the social learning problem with negative network externality. In Part II of this two-part paper, we illustrate three specific applications of Chinese restaurant game in wireless networking, cloud computing, and online social networking. In the spectrum access problem in wireless networking, we show that the overall channel utilization can be improved by taking the negative network externality into account in secondary users' decision process. The interference from secondary users to the primary user can also be reduced through learning from the sensing results of other users. In the storage platform selection problem in cloud computing, we show that customers automatically balance the loading of two platforms according to their knowledge on the platforms' infrastructures. The average loading is optimized under the best response strategies derived in Chinese restaurant game. In the deal selection on Groupon in online social networking, we show that the phenomenon of severe quality degradation of over-promotion through deals exists under the traditional social learning strategy, but not in the proposed Chinese restaurant game where customers can find a balance among several deals by taking the negative network externality into account. Moreover, we study how a new restaurant should strategically determine its deal price and advertising effort in order to maximize the revenue. When the restaurant's quality is high, he should try to advertise his restaurant as much as possible to convince customers to come for the high quality meals, while if the restaurant's quality is low, he should avoid advertising and provide a lower deal price for attracting those customers with little knowledge on the restaurant.

\bibliography{crg}
\bibliographystyle{unsrt}

\end{document}